\documentclass[12pt]{iopart}
\usepackage{iopams}  
\usepackage{graphicx}
\usepackage{color}
\usepackage[normalem]{ulem}

\begin{document}
\title[WIMP abundance and lepton (flavour) asymmetry]{WIMP abundance and lepton (flavour) asymmetry}
\author{Maik Stuke$^{1}$, Dominik J Schwarz$^1$ and Glenn Starkman$^2$ }
\address{$^{1}$Fakult\"at f\"ur Physik, Universit\"at Bielefeld, Postfach 100131, 33501 Bielefeld, Germany, and\\
					$^{2}$ CERCA/ISO, Department of Physics, Case Western Reserve University, Cleveland,OH 44106-7079 }
\eads{\mailto{mstuke} and \mailto{dschwarz}  at \mailto{physik.uni-bielefeld.de}, and\\ \mailto{glenn.starkman} at \mailto{case.edu}}
\begin{abstract}
 We investigate how large lepton asymmetries affect the evolution of the early universe at times before big bang nucleosynthesis
and in particular how they influence the relic density of WIMP dark matter. 
In comparison to the standard calculation of the relic WIMP abundance we find a decrease, depending on the lepton flavour asymmetry. We find an effect of up to 20 per cent for lepton flavour asymmetries $l_f= {\cal O}(0.1)$.
\end{abstract} 
\vspace{2pc}
\section{Introduction}

A wide variety of observations point to the existence of dark matter. Several different candidate 
particles have been proposed with masses varying from $10^{-6}$eV for Axions to $10^{16}$GeV 
WIMPzillas, as well as much more massive composite objects. Maybe the best-motivated candidate is the 
Weakly Interacting Massive Particle (WIMP) with a mass $m_{\chi}={\cal O}(\rm{GeV -TeV})$. One reason is 
that the so-called hierarchy problem of the standard model suggests the existence of an additional particle 
with a mass around the weak scale. Such a weak scale particle, if produced thermally, would have an 
abundance today, similar to the measured dark matter density \cite{Lee:1977ua}. The relic abundance 
after statistical\footnote{We distinguish here between statistical (chemical) freeze-out of 
number-changing reactions, and kinetic (thermal) freeze-out.}
freeze-out depends on the particle mass, the freeze-out temperature, and the annihilation cross section. The 
freeze-out temperature is related to the particle mass, $T_{\rm{fo}}\simeq m_{\chi}/25$. 
Below this temperature, WIMP production ceases and the WIMPS react only kinetically 
with the remaining Standard-Model particles. The kinetic freeze-out is delayed and happens at
a temperature of a few MeV \cite{Green:2005fa}. 

Assuming that all dark matter is made of WIMPS, without asymmetry between WIMPs and anti-WIMPs,
and that there are no complications such as co-annihilation, one can use the WMAP 7 yr constraint on 
the relative relic abundance 
\cite{WMAP7yr}
\begin{equation}
\Omega_{\rm{DM}}h^2 =0.1109 \pm 0.0056
\end{equation}
to fix the annihilation cross-section as a function of WIMP mass $m_{\chi}$. Several observations of candidate 
dark-matter annihilation products, like $e^{\pm}$, (anti-)protons, (anti-)deuterons, (anti-)neutrinos and 
photons constrain the dark matter cross-sections. For a recent overview see for example 
\cite{Catena:2009tm}. The standard calculations, as described above, assume no asymmetry between neutrinos and anti-neutrinos. 

In this paper, we investigate how a neutrino asymmetry affects the WIMP abundance. 
Dropping the assumption of equal numbers of neutrinos and anti-neutrinos leads to 
additional contribution to the total energy density in the early universe and thus potentially alters the WIMP 
relic abundance. This slight modification of the Standard Model, which might emerge from a fundamental 
model of neutrino masses and interactions, is well within observational limits and might play a crucial role in 
determining the relation between the relic abundance of a WIMP and its annihilation cross-section.

Very little is known about neutrino (flavour) asymmetries, due to the lack of any direct observation of cosmological neutrinos and the large number of different theories of lepto- and baryogenesis. Thus, the cosmic background of neutrinos may hide a large neutrino-anti-neutrino asymmetry, orders of magnitude larger than the baryon asymmetry \cite{Simha:2008mt}, $b\equiv\frac{n_b-n_{\bar{b}}}{s}\simeq {\cal O}(10^{-10})$.
Here $n_b$ and $n_{\bar{b}}$ denote the number densities of baryons and anti-baryons, 
respectively, while $s$ is the entropy density.   

All experimental bounds on the neutrino (flavour) asymmetries are inferred indirectly from measurements of 
the primordial abundances of light elements, the cosmic microwave background (CMB), and large scale structure (LSS) and combinations of them. 

When the universe was roughly $0.1$ seconds old and had a temperature of 
a few MeV, the neutrinos were in thermal equilibrium with the photons and $e^{\pm}$ via processes like $e^+ + e^- \leftrightarrow \nu_i + \bar{\nu}_{i}$, with $i=e$,$\mu$,$\tau$. Subsequent cooling of the universe allowed the neutrinos to kinetically decouple from the plasma. As muon and tau neutrinos interact via neutral currents only, they decouple first. The electron neutrinos interact further via charged-current interactions with the ambient electrons, neutrons and protons. For this reason, an asymmetry in the electron neutrinos influences the neutron-to-proton ratio at weak-interaction-freeze-out, and thereby  the $^4$He-abundance. Additionally any lepton flavour asymmetry could be detected through its additional contribution to the total energy density of the universe, often expressed as the number of extra effective 
relativistic degrees-of-freedom of the plasma 
\begin{equation}
\Delta N_{{\rm eff}}=\sum_{f = e,\mu,\tau}\left[\frac{30}{7}\left( \frac{\mu_{\nu_{f}}}{\pi T}\right)^2+\frac{15}{7}\left( \frac{\mu_{\nu_{f}}}{\pi T}\right)^4 \right],
\end{equation}
where $\mu_{\nu_{f}}$ are the chemical potentials of the three neutrino flavour. 

To be more specific, the neutrino flavour asymmetry influences the $^4$He abundance in two ways: 
Increasing $\mu_{\nu_e}$ would lead to a smaller $n/p$ fraction at the freeze out of the weak interaction 
rates and hence to a smaller $^4$He abundance. But this effect can be compensated by increasing 
simultaneously $|\mu_{\nu_{\mu}}|$ and/or $|\mu_{\nu_{\tau}}|$. This would raise the expansion rate and 
thus lead to a higher freeze out temperature for the weak interactions and increase the $^4$He abundance. 
Note, that the chemical potentials in $\Delta N_{\rm{eff}}$ enter quadratically and thus a 
measurment of $\Delta N_{\rm{eff}}$ does not constrain the sign of the individual flavour asymmetries. 
Thus even a vanishing net lepton asymmetry might give rise to interesting and non-vanishing lepton-flavour 
asymmetries, i.e. $l=\sum_f l_f=0$. The neutrino 
flavour asymmetries can be played against each other to match the observational bounds on light element 
abundances \cite{Olive:1991ru}.

After the epoch of Big Bang Nucleosynthesis (BBN), the total energy density of the universe consisted of photons 
and the three types of relativistic neutrinos. Altering that energy density with a large $\Delta N_{{\rm eff}}$, 
e.g. large $\mu_{\nu_f}/T$,  would affect the CMB spectrum by changing the redshift of matter 
radiation equality and by changing the amount of anisotropic stress \cite{Hou:2011ec}. 
The power spectrum of density fluctuations on small scales would be suppressed, leading to observable 
effects in the large scale structure \cite{Hou:2011ec}. 

The BBN bounds on the pseudo-chemical potentials  
$\xi_f := \mu_{\nu_f}/T$ and on  $\Delta N_{\rm{eff}}$ depend on assumptions regarding the efficiency of 
neutrino flavour equilibration via neutrino oscillations. 
Before the onset of neutrino flavour oscillations (i.e. at $T>T_{\rm{osc}}$), each individual lepton flavour,
\begin{equation}
l_f\equiv\frac{n_f-n_{\bar{f}}+n_{\nu_f}-n_{\nu_{\bar{f}}}}{s},
\end{equation} 
is conserved.
After the onset of neutrino flavour oscillations, only the total lepton number $l\equiv\sum_{f=e,\mu,\tau}l_f$ 
remains conserved, 
while the individual flavour asymmetries vary. 
It might be that $\nu$-oscillations ensure the full equilibration of three initially different flavour asymmetries 
\cite{Dolgov:2002ab, Abazajian:2002qx, Wong:2002fa, Mangano:2011ip}, 
such that the asymmetry in the electron neutrinos measured today, would be the same as in the muon and 
tau type. 
In \cite{Dolgov:2002ab,Wong:2002fa}, it is shown that even large primordial asymmetries with 
$\mu_{\nu_e}/T=0$, $\mu_{\nu_\mu}/T=-0.1$, and $\mu_{\nu_\tau}/T=0.1$ equilibrate at roughly 
$T\simeq 4$ MeV. 
Thus, one can assume that, at least for $\mu_{\nu_f}/T = {\cal O}(0.1)$,
neutrino flavour equilibration happens at $T \sim 10$ MeV, well before the onset of BBN \cite{Mangano:2011ip}.
As a consequence LSS, CMB, and BBN are blind to the earlier differences in the individual lepton flavour. 
For this case ($l = \sum_f l_f = 3 l_e$), the observation of 
primordial abundances and WMAP data constrain all three pseudo-chemical potentials:
$\xi_f \leq 0.023\pm 0.041$ \cite{Simha:2008mt},
assuming the lepton asymmetry to be constant between nucleosynthesis and photon decoupling.
The recently reported He$^4$ abundance by the ACT collaboration \cite{Dunkley:2010ge} and the WMAP7 data release lead to a significantly larger range 
$-0.14 \leq \xi_f \leq 0.12$ \cite{Krauss:2010xg}. 

Considering only partial equilibration of three initially different flavour asymmetries via oscillations 
(i.e.~$\mu_{\nu e} \neq \mu_{\nu_\mu} \neq \mu_{\nu_\tau}$), weakens the bounds on $\xi_f$. 
Assuming $\mu_{\nu_e} \ll \mu_{\nu_f}$ and an effective number of neutrinos 
$N_{\rm eff}=3.3^{+0.7}_{-0.6}$ while neglecting flavour equilibration, 
leads to the bound $|\xi_f|\leq 2.34$ (here $f = \mu,\tau$) 
and a total lepton asymmetry of $|l| \leq 5$ \cite{Simha:2008mt}. We conclude, large lepton asymmetries 
and large lepton flavour asymmetries before the onset of neutrino oscillations are compatible with current observations. 

The consequences of lepton asymmetries on the cosmic QCD 
transition at $T_{\rm qcd} \sim 200$ MeV are discussed in \cite{Schwarz:2009ii}.
There it is shown that large neutrino asymmetries influence 
significantly the dynamics of the QCD transition. As the electric charge, baryon and lepton 
flavour are preserved at the same time, a non-negligible lepton asymmetry can induce a non-vanishing 
baryon-chemical potential, which in turn influences the pseudo-critical temperature or even turns a crossover transition into a first order transition.   

In this work we investigate an even earlier event: statistical 
freeze-out of the abundance of weakly interacting dark matter. 
We show that large lepton (flavour-) asymmetries ($|l| > 0.01$) give rise to sizeable effects on the relic abundance of WIMP dark matter.

In the following chapter we we will give an (incomplete) overview of arguments how large lepton (flavour) asymmetries can be generated. In the third chapter we show then, how large lepton (flavour) asymmetries influences the thermodynamic description of the early universe. We give analytic estimations and numerical results and show how these effect the relative relic abundance of WIMP dark matter in the fourth chapter before we conclude our work in chapter 5.
Throughout the article we set $c=\hbar=k_B=1$.

%
\section{A short review of models with large lepton (flavour) asymmetry} 
%
Several ideas of baryo- and leptogenesis have been proposed 
that lead to a universe with large lepton flavour asymmetries but small baryon asymmetry.
For example, in grand unified theories asymmetries $|{l}|\gg b$ can be realized based on gauge invariant initial particle asymmetries \cite{Harvey:1990qw}.
Another mechanism is provided by supersymmetric theories, c.f.~\cite{McDonald:1999in}, 
which have the intriguing feature of flat directions. Made up of squark or slepton fields, 
these directions can carry baryon and/or lepton number. During cosmological inflation these 
squark and sleptons are free to fluctuate and to form scalar condensates, carrying baryon 
and/or lepton number, and to release these charges by decaying  to Standard-Model particles. 
This is called the Affleck-Dine mechanism \cite{Affleck:1984fy,Dolgov:1990zm}, 
and in principle can give rise to large asymmetries in baryons or leptons.  

In the early  universe, large asymmetries have to be produced after an inflationary phase, 
otherwise they diluted away by the inflationary expansion and washed out 
 by the huge increase in entropy density during post-inflationary reheating. 
Another possibility to dilute a large lepton (flavour) asymmetry are sphaleron transitions at high temperatures. These transitions violate $b$ and $l$, but preserve $b-l$. Sphalerons are the reason for suggesting ${\cal O}(b)\simeq {\cal O}(l)$. Note that this does not necessarily mean 
${\cal O}(b)\simeq {\cal O}(l_f)$. For instance, two flavours could be orders of magnitude larger, if 
their sum almost cancels, $l_i \simeq - l_j$ for lepton flavours $i\neq j$. 

However, sphaleron transitions might never be in equilibrium 
and the argument in favor of $l = {\cal O}(b)$ would disappear. 
It was shown that, if the total asymmetry $\sum_f l_f$ is larger than a critical value 
$l_{\rm c}\simeq 10^{-2}$ then electroweak symmetry is never restored and sphalerons are suppressed for 
all times \cite{Linde:1976kh,McDonald:1999nv}. A large lepton (flavour) asymmetry would survive
until today. It is remarkable that recent bounds on the electron-neutrino asymmetry from WMAP-7yr 
alone and combined with the ACT data are in the vicinity of this limit.
 
In \cite{Casas:1997gx} the authors combine the Affleck-Dine mechanism with the suppressed sphalerons. They consider a model with a sneutrino condensate to generate large $l$-asymmetries, $l>l_{\rm c}$, but no baryon asymmetry. In their specific model they take the minimal supersymmetric Standard-Model with three right-handed neutrino singlet superfields. For a mass of the lightest neutrino of ${\cal O}(10^{-4} {\rm eV})$ and a preferred mass of the lightest 
right-handed neutrino of ${\cal O}({\rm TeV})$, 
their model leads to asymmetries $l\simeq 10^{-2}$ to 1. 
In \cite{Liu:1993am,McDonald:1999nv} it is shown how the electroweak symmetry non-restoration due to a 
large $l$-asymmetry works with a small baryon asymmetry. 
How to regulate the baryon asymmetry in Affleck-Dine mechanisms is discussed in \cite{Campbell:1998yi}. 
An interesting model is also discussed in \cite{MarchRussell:1999ig}, where $l_e=-l_{\mu}$ and $l_{\tau}=0$.

%
\section{Thermodynamics in the early universe}
%

It is an excellent approximation to assume that entropy is conserved in the early Universe, even through several phase transitions \cite{BdVS}. 
Even if the QCD transition were of first order, the amount of entropy produced would be tiny \cite{SSW,BdVS}. 
It is convenient to work with specific densities of conserved quantities.  
The specific baryon asymmetry $b$ is related to the baryon number density $n_B$, via $b = n_B/s$, 
where $s$ denotes the entropy density. 
Constraints from BBN and CMB are often formulated in terms of $\eta_B = n_B/n_\gamma = b s/n_\gamma$. 
Note that $\eta_B(t_{\rm bbn}) \neq \eta_B(t_{\rm cmb})$, while $b$ is constant. 

Before neutrino decoupling, the radiation fluid of the early Universe is a tightly coupled plasma 
in thermal and statistical equilibrium with respect to strong, electromagnetic and weak forces. 
It is characterized by a single temperature $T$ and chemical potentials $\mu_i$ for each particle 
species (with mass $m_i$) of the Standard-Model 
($i=$ quarks/hadrons, leptons and gauge bosons and we count particles and antiparticles separately). 
The chemical potentials of anti-particles $\mu_{\bar{i}}=-\mu_i$, as long 
as particles are relativistic ($T > m_i/3$). (The chemical potentials of the non-relativistic particles (e.g. baryons) are 
irrelevant, only because they are apparently too small for these species  to contribute appreciably  to the energy density, pressure and entropy of the Universe  
during the epochs in question.)

In this work we consider temperatures between the electroweak and the QCD transition, 
$T_{\rm ew}>T>T_{\rm qcd}$, where $T_{\rm ew}\sim 200$ GeV and $T_{\rm qcd}\sim 200$ MeV.
For WIMP masses between about $5$ GeV and $4$ TeV,  it is within this temperature range 
that  WIMP annihilation freezes-out and the WIMP number density is determined.

The equilibrium distribution function $f_i$ of a  particle $i$ and its anti-particle $\bar{i}$ is given for 
bosons ($-$) and fermions ($+$) by the Bose-Einstein and Fermi-Dirac distributions, 
\begin{eqnarray}
  f_i (E) = \frac{1}{{\rm exp}\frac{E-\mu_i}{T} \mp 1}, \quad
  f_{\bar{i}}(E) = \frac{1}{{\rm exp}\frac{E+\mu_i}{T} \mp 1},
\end{eqnarray}
where $E$ denotes the energy of the particles. 
We are interested in the difference of the number densities of particles and their anti-particles, 
i.e.~their net number density,
\begin{eqnarray}
\label{nint}  n_i = \frac{g_i}{2\pi^2} \int^{\infty}_{m_i}{E(E^2-m_i^2)^{1/2}\left[f_i(E) - f_{\bar{i}}(E) \right]{\rm d}E},
\end{eqnarray}
where $g_i (= g_{\bar i})$ denotes the number of helicity degrees of freedom of a particle
species (e.g.~$g_e = 2$). For ultra-relativistic fermions the net number density simply becomes
\begin{equation}
n_i = \frac{g_i}{6} \mu_i T^2 \left[ 1 + \left(\frac{\mu_i}{\pi T}\right)^2 \right].
\end{equation}   

All particles and anti-particles contribute to the energy density of the Universe, which is given by
\begin{eqnarray}\label{intg}
\epsilon = \sum_{j = i,\bar{i}} \frac{g_j}{2\pi^2}\int^{\infty}_{m_j}{E^2(E^2-m_j^2)^{1/2}f_j(E){\rm d}E},
\end{eqnarray}
where the index $j$ can stand for any particle or anti-particle species.

In equilibrium, electromagnetic and weak interactions provide a set of relations between the chemical potentials. It follows that all gauge bosons have vanishing chemical potentials, and for relativistic fermions
\begin{eqnarray}
\mu_i = - \mu_{\bar i},\\
\mu_u + \mu_f = \mu_d + \mu_{\nu_f},\\
\mu_u = \mu_c = \mu_t, \\
\mu_d = \mu_s = \mu_b.
\end{eqnarray}
The indices $u$, $d$, $c$, $s$, $t$ and $b$ represent the up, down, charm, strange, top and bottom 
quarks. Lepton flavour ($e$, $\mu$, or $\tau$) is denoted by $f$. 

This leaves us with five independent chemical potentials (i.e.~three for the neutrino flavours and two 
for the up and down quarks) plus the temperature. These six variables are uniquely determined by 
five conservation laws and the Friedman equations. The five conserved quantities in the energy range 
of interest (between $T_{\rm ew}$ and $T_{\rm qcd}$) are electric charge ($q$), 
baryon number and three lepton flavour numbers. 
\begin{eqnarray} 
\label{charge}s\, q    &=& - \sum_{i=e,\mu,\tau} n_i + \frac 23 \sum_{i= u,c,t} n_i - \frac 13 \sum_{i = d,s,b} n_i \\
\label{baryon}s\,  b   &=& \frac 13 \sum_{i= u,d,c,s,t,b} n_i \\
\label{lflavour}s\,  l_f &=& n_f + n_{\nu_f}, \quad f = e, \mu, \tau 
\end{eqnarray} 
Lepton flavour is conserved, since the timescale for $\nu$-oscillations is much larger then the 
Hubble time for $T>$ few MeV. Additionally we assume that these global conservation laws hold 
locally, i.e.~there are no electric currents and no baryon or lepton (flavour) diffusion. 

We assume a charge neutral universe ($q=0$) and fix the baryon asymmetry to
$b = 9 \times 10^{-11}$ \cite{Simha:2008mt}. While these are well established assumptions, measurements of the lepton flavour asymmetries in the early Universe are not available. 
We thus keep them as free parameters. In \cite{Schwarz:2009ii} it is shown in 
detail how to extract all chemical potentials $\mu_i(T;b,\{l_f\})$, including details on the numerical 
method used to solve the five conservation equations as a function of $T$. 

%
\subsection{Effective relativistic degrees of freedom} 

Let us now take a closer look at the contribution of lepton flavour asymmetries on the effective 
relativistic degrees of freedom contributing to the total energy density 
\begin{equation}
g_{\ast}(T,\{\mu_i\}){\equiv}\frac{30}{\pi^2T^4} \epsilon(T,\{\mu_i\}).
\end{equation}
Together with the solution $\mu_i = \mu_i(T;b,l_e, l_\mu, l_\tau)$ we find
\begin{equation}
g_{\ast}=g_{\ast}(T;b,l_e,l_{\mu},l_{\tau}).
\end{equation}
In the ultra-relativistic case ($m_i=0$) and for vanishing baryon and lepton asymmetry (i.e.~vanishing chemical potentials) we recover
\begin{equation}
g_{\ast}(T,\{0\})=\sum_{i = {\rm bosons}} g_i + \frac{7}{8}\sum_{i={\rm fermions}} g_i.
\end{equation}

Large lepton flavour asymmetries generically lead to large chemical potentials of all fermion species. 
For ultra-relativistic fermions, (\ref{intg}) can be solved exactly. Assuming $m_i = 0$ and 
$\mu_i=-\mu_{\bar{i}}$ leads for the energy density of a fermion species to
\begin{equation}
\epsilon_{i + \bar{i}} \stackrel{\mu_i=-\mu_{\bar{i}}}{=}  \frac{\pi^2}{30} T^4 (2 g_i)
\left[\frac{7}{8} + \frac{15}{4} \left(\frac{\mu_i}{\pi T}\right)^2 
+ \frac{15}{8} \left(\frac{\mu_i}{\pi T}\right)^4 \right].
\end{equation}
This leads to an increase of $g_\ast$, due to non-vanishing chemical potentials, 
\begin{eqnarray}
\label{gast}
g_{\ast} (T,\{\mu_i\}) 
&=g_{\ast}(T,{0}) + \Delta g_{\ast}(T,\{\mu_i\}),
\end{eqnarray}
with
\begin{equation}\label{Deltagast}
\Delta g_{\ast} (T,\{\mu_i\}) =
\sum_i g_i \left[\frac{15}{4} \left(\frac{\mu_i}{\pi T}\right)^2 + 
          \frac{15}{8} \left( \frac{\mu_i}{\pi T}\right)^4 \right]. 
\end{equation}
Any nonzero $\Delta g_{\ast}$ would therefore increase the total energy density and thus the Hubble  expansion rate. 
This increased expansion rate would alter the relic abundances of light elements, and so must be checked against observations.
The increased abundance of relativistic species is normally expressed as an increase of the effective number of neutrinos $N_{\nu_{\rm{eff}}}$.
>From the LEP measurement of the decay width of the $\rm{Z}^0$ boson, we expect three active neutrino species with masses well below the electroweak scale, thus
\begin{equation}
N_{\nu_{\rm{eff}}}= 3 + \frac{30}{7}\sum_{f}\left[\left(\frac{\mu_{\nu_f}}{\pi T}\right)^2 + 
    \frac{1}{2}\left(\frac{\mu_{\nu_f}}{\pi T} \right)^4\right].
\end{equation}

We also have to take a closer look at the $\mu$-dependence of $s/T^3$. In kinetic theory, including quantum statistics, the entropy density in terms of the distribution function $f$ is \cite{Bernstein}
\begin{equation}\label{sint}
s = -\int{\left[f {\rm ln}f \mp(1 \pm f){\rm ln}(1 \pm f)\right]\frac{{\rm d}^3p}{(2\pi)^3}},
\end{equation}
where upper and lower signs refer to boson and fermion statistics, respectively.

For ultra-relativistic particles, vanishing chemical potentials and all particle species at temperature 
$T$, we recover the well known result 
\begin{eqnarray}\label{entropyT}
s(T) \stackrel{m,\mu=0}{=} && \frac{2\pi^2}{45}T^3g_{\ast}.
\end{eqnarray} 
Taking chemical potentials into account leads to extra contributions
\begin{eqnarray}\label{entropymu}
s(T,\mu_i)&\stackrel{m=0}{=}& 
\frac{2\pi^2}{45} T^3 \left[g_\ast + \frac{15}{8}\sum_{i={\rm fermions}}g_i\left(\frac{\mu_i}{\pi T}\right)^2\right] \\
&=&\frac{2\pi^2}{45}T^3\left( g_{\ast}+\Delta g_{s_{\ast}}\right).
\end{eqnarray}
Note that $\Delta g_{s_{\ast}} \neq \Delta g_{\ast}$ and, in contrast to the energy density,  terms quartic in the chemical potentials do not show up in the entropy density. When $T \sim m_i$, we have to 
calculate the entropy density numerically. 

%
\subsection{Analytic estimates and numerical results}

For an analytic estimate of the effect of lepton flavour asymmetries, we neglect all masses 
of quarks and leptons and assume that all lepton flavour asymmetries are small enough to 
justify $\mu_i/(\pi T) \ll 1$. Further we assume $m_W/3  > T > m_b/3$, thus $g_\ast = 345/42$. 
{}From the conservation of charge, baryon number and lepton flavour we find 
\begin{eqnarray}
0      &=& \frac 1 3 T^2 (4 \mu_u - 3 \mu_d - \mu_e - \mu_\mu - \mu_\tau) + {\cal O}(\mu_i^3), \\
\frac{23 \pi^2}{6} T^3 b   &=& \frac{1}{3} T^2 (2 \mu_u + 3 \mu_d) +  {\cal O}(b \mu_i^2, \mu_i^3), \\
\frac{23 \pi^2}{6} T^3 l_f &=& \frac 1 6 T^2 (2 \mu_f + \mu_{\nu_f}) +  {\cal O}(l_f \mu_i^2,\mu_i^3).
\end{eqnarray} 
Solving this set of equations results in 
\begin{eqnarray} 
\label{mudApr} 
\frac{\mu_d}{\pi T} &=& \pi \left[\frac{5}{2} b - \frac{2}{3} l \right], \\
\frac{\mu_u}{\pi T} &=& \pi \left[2 b +  l \right], \\
\frac{\mu_f}{\pi T} &=& \pi \left[\frac{1}{6} b - \frac{5}{9} l + \frac{23}{3} l_f \right], \\
\label{munuApr} 
\frac{\mu_{\nu_f}}{\pi T} &=& \pi \left[- \frac{1}{3} b + \frac{10}{9} l + \frac{23}{3} l_f \right].
\end{eqnarray}
Lepton (flavour) densities large compared to the baryon density affect not only the number densities of leptonic species, but also those of quarks.  This can lead to increase of the 
effective degrees of freedom in the energy density and the entropy density. Note that all relativistic particle species equipped with baryon or lepton number contribute to this effect, as long as they are in statistical equilibrium. 

\subsubsection{Flavour symmetric lepton asymmetry}

Let us first assume that all lepton flavour numbers are the same, $l_e = l_\mu = l_\tau = l/3$. We also 
assume $b \ll |l| \ll 1$. Thus we can put $b=0$. This results in
\begin{equation} 
\frac{\mu_d}{\pi T} = - \frac{2\pi}{3} l, \quad
\frac{\mu_u}{\pi T} = \pi l, \quad
\frac{\mu_f}{\pi T} = 2\pi l, \quad
\frac{\mu_{\nu_f}}{\pi T} = \frac{11\pi}{3}l,
\end{equation}
and allows us to estimate the change in the effective degrees of freedom, 
\begin{equation}\label{gast2}
\Delta g_{\ast}(T,b=0,l_f = l/3) = \frac{1265}{2} \pi^2 l^2 \approx 6.2 \times 10^3 l^2.
\end{equation}  
Thus $\Delta g_{\ast}/g_{\ast} \approx 760 l^2$, which we assumed to be small for the purpose of the analytic approximation, i.e.~$l < 10^{-2}$.    

For lepton asymmetries $l > 10^{-2}$ we rely on a numerical solution of the five conservation equations and include all particles from the Standard-Model of particle physics with their measured physical 
masses (the unknown masses of the Higgs and the neutrinos are irrelevant in the regime of interest). 
We solved the equations (\ref{charge}) to (\ref{lflavour}) using (\ref{nint}) and (\ref{sint}) with the method described in detail in  \cite{Schwarz:2009ii}.

\begin{figure}[htb]
	\centering
		\includegraphics[angle=270,width=0.80\textwidth]{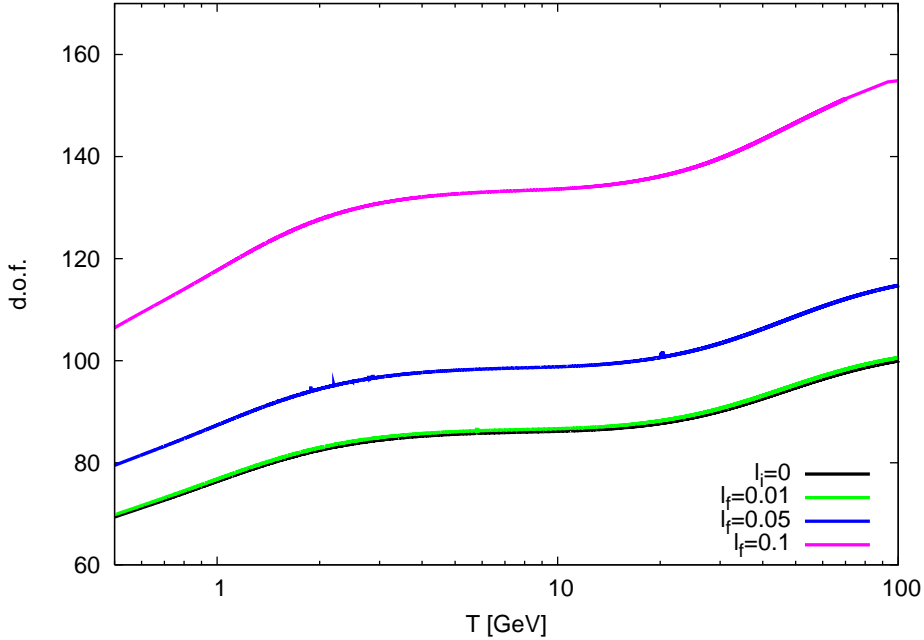}
	\caption{The numerical solution for the flavour symmetric case $l=3l_f$. The effective degrees of freedom of all particles in statistical equilibrium versus the temperature in GeV on logarithmic scale. The black line corresponds to standard case, where lepton asymmetries are neglected. The blue line shows the influence of $l_f=0.01$, the green  $l_f=0.05$ and the red $l_f=0.1$.}
	\label{fig:doflsym}
\end{figure}

The numerical results for the effective helicity degrees of are shown in Fig.~\ref{fig:doflsym}. We found that an asymmetry $l_f=0.01$ leads to a small deviation from the standard case with $b=l=l_f=0$. If we apply the experimentally given upper bound for the electron neutrino asymmetry to all flavour, we found for $l_f=0.1$ approximately additional 50 degrees of freedom in the early universe between $1<T<50$ GeV. 
 
\subsubsection{Flavour asymmetric lepton asymmetry}

\begin{figure}
	\centering
		\includegraphics[angle=270,width=0.80\textwidth]{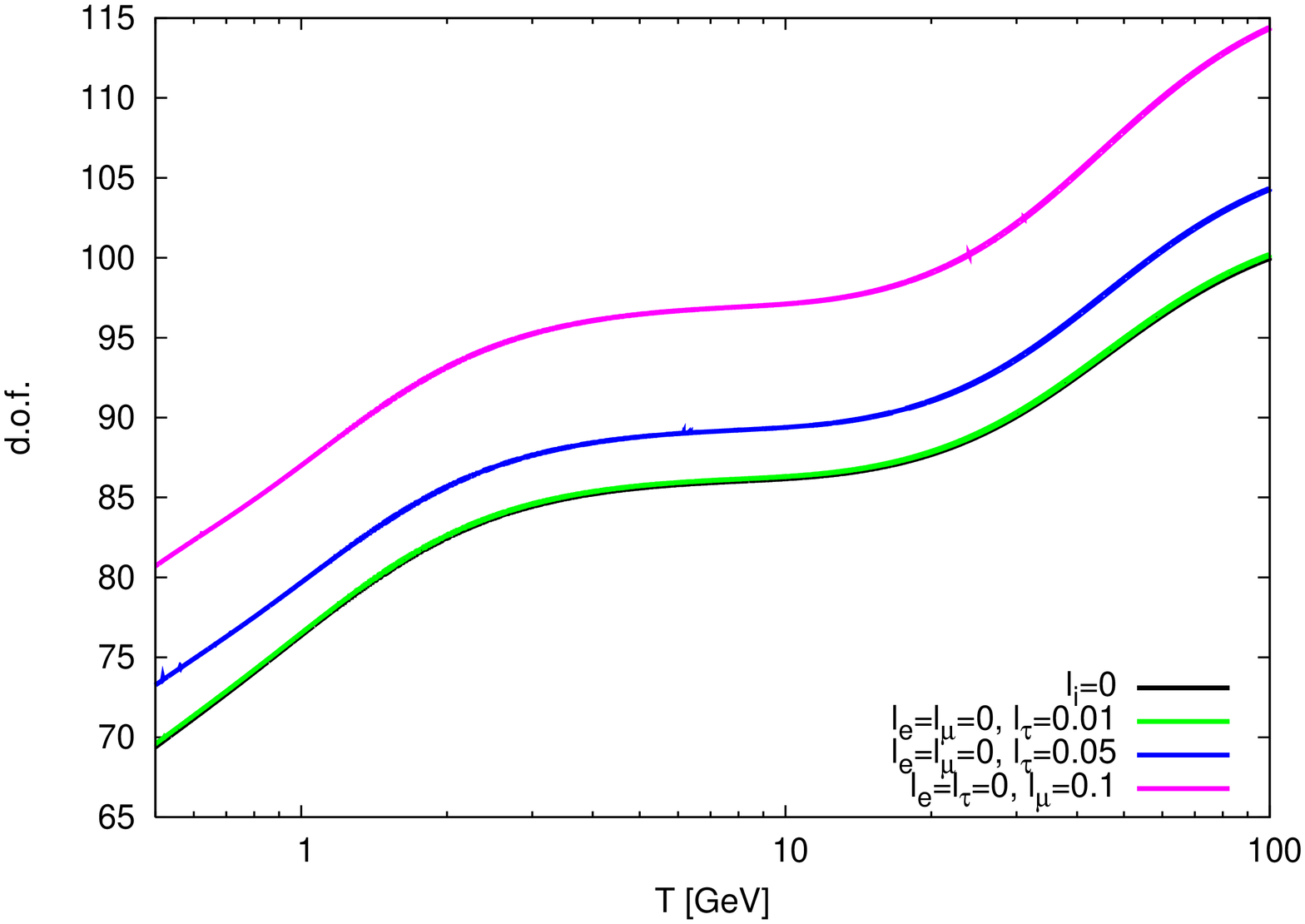}
	\caption{The numerical solution for the flavour asymmetry $l=l_{\tau}$. The effective degrees of freedom of all particles in statistical equilibrium versus the temperature in GeV on logarithmic scale. The black line corresponds to standard case, where lepton flavour asymmetries are neglected. The green line shows the influence of $l_{\tau}=0.01$, the blue stands for $l_{\tau}=0.05$ and the red for $l_{\tau}=0.1$.}
	\label{fig:dofl=ltau}
\end{figure}

Let us now have a closer look at scenarios in which at least one of the three flavour lepton numbers satisfies $|l_f| \gg b$, but we restrict to $|l_f| \leq 1$ for all flavour. For simplicity we can put $b=0$. The first interesting situation is that one flavour asymmetry dominates, say $l_\tau \neq 0$ and the other flavour asymmetries vanish. In that case we would find that the quark chemical potentials are affected:
\begin{eqnarray} 
\frac{\mu_d}{\pi T} &=& - \frac{2\pi}{3} l_\tau, \\
\frac{\mu_u}{\pi T} &=& \pi l_\tau, \\
\frac{\mu_{e,\mu}}{T} &=&  - \frac{5 \pi}{9} l_\tau, \\
\frac{\mu_\tau}{T} &=& \frac{64 \pi}{9} l_\tau, \\
\frac{\mu_{\nu_{e,\mu}}}{T} &=& \frac{10 \pi}{9} l_\tau, \\
\frac{\mu_{\nu_\tau}}{T} &=&  \frac{79 \pi}{9} l_\tau.
\end{eqnarray}
The numerical results for this situation are presented in Fig.~\ref{fig:dofl=ltau}. We see again a tiny deviation from the standard case for $l_{\tau}=0.01$. For $l_{\tau}=0.1$ there would be around 10 more degrees of freedom. The effect is smaller compared to symmetric case since the total $l$ is smaller.

\begin{figure}[htbp]
	\centering
		\includegraphics[angle=270,width=0.80\textwidth]{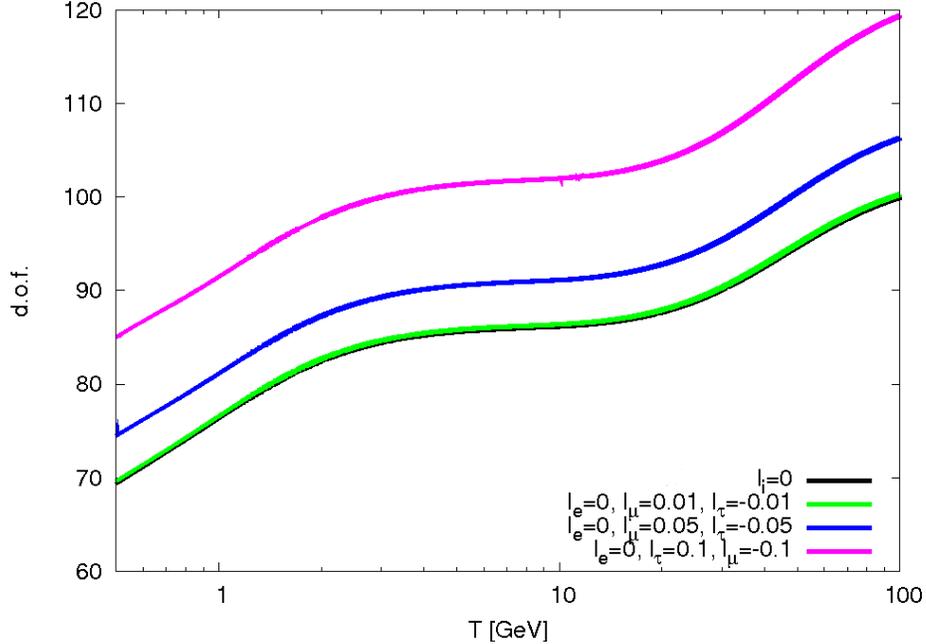}
	\caption{The numerical solution for the flavour asymmetry $l_{\mu}=-l_{\tau}$ and $l=l_e=0$. The effective degrees of freedom of all particles in statistical equilibrium versus the temperature in GeV on logarithmic scale. The black line corresponds to standard case, where lepton flavour asymmetries are neglected. The green line shows the influence of $l_{\tau}=0.01$, the blue stands for $l_{\tau}=0.05$ and the red for $l_{\tau}=0.1$.}
	\label{fig:doflmu=ltau2}
\end{figure}

Another interesting scenario is a vanishing total lepton asymmetry $l=0$, but $l_{\mu}=-l_{\tau} \neq 0$. In this case quark chemical potentials would not be affected:
\begin{equation} 
\mu_d = \mu_u = \mu_e = \mu_{\nu_e} = 0, \quad
\frac{\mu_f}{\pi T} = \frac{\mu_{\nu_f}}{\pi T}  = \frac{23 \pi }{3} l_f, \quad
f = \mu,\tau.
\end{equation}
We find $\Delta g_\ast = 5 (23 \pi)^2 l_\tau^2 \approx 2.6 \times 10^4 l_\tau^2$. As we assumed 
for the analytic approximation that the modification is small, its regime of validity is limited to 
$|l_\tau| < 10^{-2}$. 
The numerical results for vanishing $l$, but non-vanishing lepton flavour asymmetry 
are presented in Fig.~\ref{fig:doflmu=ltau2}. In the calculations for the degrees of freedom the sign of a possible asymmetry does not play any role, since they enter squared. For $l_{\mu}=-l_{\tau}=0.1$ additional 20 degrees of freedom appear. What makes this scenario the most interesting, is the possibility for even larger asymmetries. If we assume $l_{\mu}=-l_{\tau}=1$ we find more then 600 additional degrees of freedom.

%
\section{Effect on decoupling of WIMP dark matter}
%
%
The WIMP is a very well-motivated candidate to solve the dark matter puzzle \cite{Jungman:1995df}. 
In the following we assume the WIMP to be the single component dark matter particle, without asymmetry between WIMPs and anti-WIMP. 
In the hot early universe these particles with masses typically between 10 and 1000 GeV are in thermal and statistical  equilibrium with the radiation content. Their statistical freeze-out, when WIMPs decouple statistically from the radiation plasma, happens at $T_{\rm fo} \simeq m/25$, leading to a mass dependent interval of 0.4 GeV $<T_{\rm fo}<$40 GeV (see e.g.~\cite{Green:2005fa}). In the following we want to investigate the effect of changing the Standard Model boundary condition by allowing large neutrino asymmetries on the WIMP freeze-out.

To calculate the relic density, one assumes annihilations of the WIMPs $X\bar X\rightarrow \cdots$ with a typical weak interaction cross section $\sigma \propto G_{\rm F}^2$. The corresponding 
Boltzmann equation then leads to an equation for their net particle density 
\begin{equation}
\dot{n} + 3Hn = -\left\langle \sigma |v| \right\rangle (n^2-n_{\rm eq}^2),
\end{equation}  
with the Hubble parameter $H$, the total annihilation cross section $\sigma$ and the relative velocity 
of the annihilating particles $v$. The index ``eq'' indicates the assumption of 
thermal and statistical equilibrium.
Since the WIMP is non relativistic at the time of decoupling, the annihilation cross section can be 
approximated for most cases as $\langle \sigma v \rangle= 
a + b/x + {\cal O}(x^{-2})$, 
where the numbers $a$ and $b$ describe s- and 
p-wave annihilation and $x \equiv m/T$, for more details see e.g.~\cite{Jungman:1995df}.   
It is common to introduce the specific WIMP abundance $Y = n/s$ and to assume the conservation 
of entropy:
It is convenient to analyze the freeze-out process in a comoving volume, thus we introduce the specific WIMP abundance $Y = n/s$ and $Y_{\rm eq} = n_{\rm eq}/s$ and the ratio $x \equiv m/T$:
\begin{equation}
\frac{{\rm d} Y}{{\rm d}x}=-\frac{\langle \sigma |v| \rangle s}{3c_s^2 H x}(Y^2 - Y_{\rm eq}^2).
\end{equation}
In radiation domination $3 c_s^2 = 1$ is 
a good approximation.
As freeze-out happens when the rate of annihilations drops below the Hubble rate, we  find 
$Y_{\infty} \approx Y_{\rm fo} \approx [H/(\langle \sigma v\rangle s)]_{\rm fo}$. 
Thus an approximate solution for the WIMP relic abundance today can be derived as
\begin{equation}
Y_{\infty}\equiv Y(x\rightarrow \infty) \approx \left(\frac{\sqrt{g_\ast}}{g_{S\ast}}\right)_{\rm fo} 
\frac{1}{m_{\rm Pl}  m (a/x_{\rm{fo}} + b/x^2_{\rm{fo}})}.
\end{equation}
for more details see e.g.~\cite{Lee:1977ua,Drees:2007kk}.

It is convenient to express the relative WIMP mass density 
$\Omega_{\rm wimp} = \rho_{\rm wimp}/\rho_{\rm c}$ with the critical density $\rho_{\rm c}$. 
The present relic abundance is then $\rho_{\rm wimp} =
m  n_{0}= m s_0Y_{\infty}$, where the subscript zero denotes the value today:
\begin{eqnarray}\label{Omegachi}
\Omega_{\rm wimp}h^2 &=& 
2.7 \times 10^{10}\frac{m}{100 \rm{GeV}} Y_{\infty} \\
 &\simeq& \frac{8.5 \times 10^{-11}}{\rm{GeV}^2} \left(\frac{\sqrt{g_\ast}}{g_{S\ast}}\right)_{\rm fo} 
\frac{x_{\rm{fo}}}{a + b/x_{\rm{fo}}}.
\end{eqnarray}
The abundance of a WIMP particle is inverse proportionally to its annihilation cross section, a more strongly interacting particle stays longer in equilibrium. The dependence on helicity degrees of freedom is apparent in the denominator, but there is also an implicit dependence in 
$x_{\rm fo} \approx c -  \ln(g_{\ast})$, where the constant c includes a logarithmic dependence 
on the wimp mass, its cross section and its helicity degrees of freedom \cite{Green:2005fa}. 

As long as $\Delta g_\ast \ll g_\ast$, which is the case for $l_f \ll 1$, we find
\begin{equation}
\label{Omega}\frac{\Delta \Omega_{\rm wimp}}{\Omega_{\rm wimp}} = 
\frac 12 \left( 1 - (1 + \ell) \frac 1{x_{\rm fo}} \right) 
\frac{\Delta g_\ast}{g_\ast} - \frac{\Delta g_{S\ast}}{g_{S\ast}}
\end{equation}
where $\ell = 0,1$ when the s- or p-wave annihilation channel dominates. This can be further simplified using the expressions obtained for the ultrarelativistic degrees of freedom. In this case the contributions from $\mu^2$ terms in the first two terms cancel out, 
\begin{equation}
\frac{\Delta \Omega_{\rm wimp}}{\Omega_{\rm wimp}} \sim
- (1 + \ell) \frac 1{x_{\rm fo}} \frac{15}{8} \sum_i g_i \left(
\frac{\mu_i}{\pi T}\right)^2.  
\end{equation}
However, this will not be the case close to mass thresholds. For the flavour symmetric case, with 
$x_{\rm fo} \approx 25$ and assuming p-wave annihilation, we find $\Delta  \Omega_{\rm wimp}/
\Omega_{\rm wimp} \simeq - 30 l^2$, i.e.~a reduction of the WIMP relic density by 30\% (7\%) for 
$l = 0.1 (0.05)$. However, for $l > 10^{-2}$ higher order contributions of the statistical potentials 
are important and the interesting regime must be analyzed numerically. 

\begin{figure}[htbp]
	\centering
		\includegraphics[angle=270,width=0.80\textwidth]{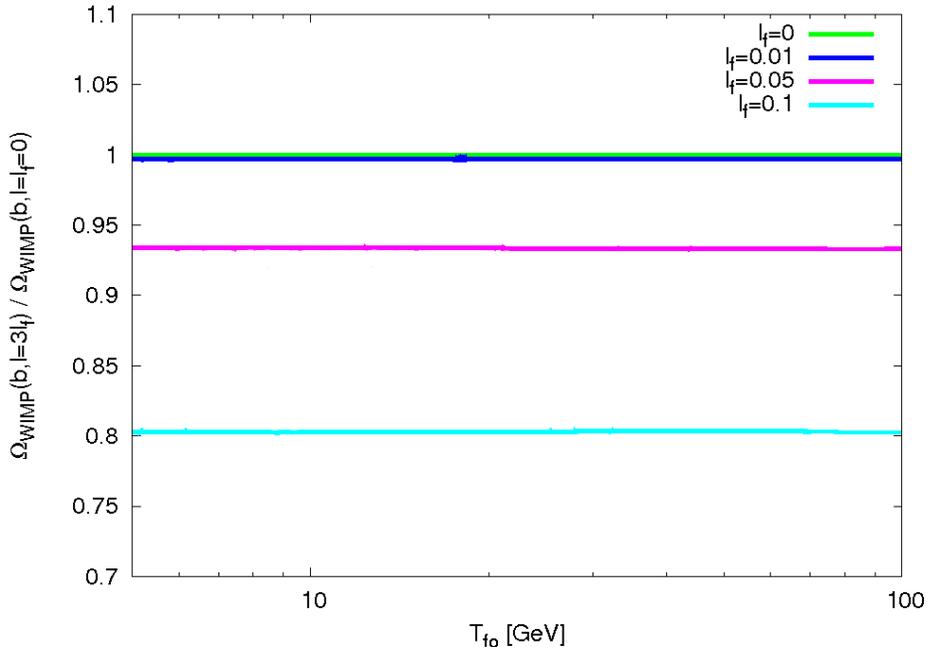}
	\caption{Comparison between the relative WIMP dark matter abundance without lepton flavour asymmetries $\Omega_{DM}(l=l_f=0)$ and several large asymmetries $l_f=l_e=l_{\mu}=l_{\tau}$, plotted versus the freeze out temperature $T_{\rm{fo}}$ on a logarithmic scale. We observe for flavour asymmetries $l_f=0.05$ an effect of approximately 7 percent. For $l_f=0.1$ this effect increases to 20 percent.}
	\label{fig:Deltamulf}
\end{figure}

For the numerical analysis we assume the governing effect in helicity degrees of freedom. The difference of the relative relic density introduced by lepton (flavour) asymmetries at a given freeze out temperature can be approximated
\begin{equation}
\frac{\Omega_{\rm wimp}(l,l_f\neq0)}{\Omega_{\rm wimp}(l,l_f=0)} \sim \left[\sqrt{\frac{g_{\ast}(l,l_f\neq0)}{g_{\ast}(l,l_f=0)}}~ \frac{g_{S\ast}(l,l_f=0)}{g_{S\ast}(l,l_f\neq0)}\right]_{x_{\rm{fo}}}.
\end{equation} 
A numerical analysis including all mass thresholds and not restricted to small chemical potentials 
is shown in figure \ref{fig:Deltamulf} for different lepton asymmetries $l_f$.
We plot the ratio $\Omega_{\rm{wimp}}(l,l_f)/\Omega_{\rm{wimp}}(l=l_f=0)$ as a function of 
WIMP freeze-out temperature. 
In the flavour symmetric case (figure~\ref{fig:Deltamulf}) we find 
that the analytic estimate for lepton asymmetries overestimates the effect.
We observe an effect of order 1 $\%$ for $l_f=0.01$ and of almost $20\%$ for $l_f=0.1$.

%
%
\section{Conclusion}
%
%
%

In this work we extend the study of cosmic lepton and lepton flavour asymmetries to the early Universe 
before the cosmic QCD transition, a subject studied previously \cite{Schwarz:2009ii}.

The standard model of cosmology does not know any lepton or baryon number violating processes after 
the electroweak phase transition. A lepton flavour asymmetry produced before or during the transition 
would remain constant until the onset of neutrino flavour oscillations at $T\simeq {\cal O}(10)$ MeV. Neutrino oscillations ensure that all lepton flavour asymmetries
agree at the time of BBN, i.e.~$l_e=l_{\mu}=l_{\tau}$,  
and thus bounds from BBN and CMB observations apply to the total lepton asymmetry $l$ only.  
These bounds are rather weak and allow $|l| \gg b$. 

Before the onset of neutrino oscillations, scenarios with different individual flavour asymmetries are even 
less constrained, e.g.~$l_e=b$ and $- l_{\mu} = l_{\tau} \gg b$.

In this work we took a closer look on the the freeze-out of WIMP dark matter. This event is of 
particular interest, since observations of these relics would open a window to the pre-BBN era. 
We have shown, that lepton asymmetries might have a sizable effect on the relic abundance of the 
WIMP dark matter. The relative relic abundance depends on the annihilation cross 
section, the WIMP mass and lepton flavour asymmetries.
 
A reduction of the WIMP abundance due to large lepton (flavour) asymmetries happened due to
an increase of the effective relativistic degrees of freedom of the radiation plasma. Large lepton 
asymmetries lead generically to large chemical potentials of all fermion species, which in turn contribute 
to the energy and entropy densities of the universe. We presented 
analytical estimates and numerical studies of this effect. Figures 
\ref{fig:doflsym} to \ref{fig:doflmu=ltau2} show how the effective relativistic degrees of freedom 
$g_{\ast}$ increase for different distributions of individual lepton flavour asymmetries.

We demonstrated how a large $l_f$ influences the relic WIMP abundance in figure \ref{fig:Deltamulf}. 
If the asymmetries are equal in in all flavour, $l_e=l_{\mu}=l_{\tau}$ and $|l_f|=0.1$ we found a 
huge effect on the relative relic abundance of approximately 20 per cent. Even if the total 
lepton asymmetry is of the order of the baryon asymmetry, but individual flavour asymmetries  
are large before the onset of neutrino oscillations, a reduction of the WIMP relic density of up to 
$10\%$ is possible. The effect presented in this work adds to the astrophysical and particle physics 
uncertainties to be accounted for in direct and indirect dark matter searches. Large lepton
(flavour) asymmetries could be compensated by decreasing annihilation cross sections, in order 
to keep the relic abundance of a WIMP candidate fixed.

\ack 

We thank Tanmay Vachaspati and Yi-Zen Chu for discussions. M.S. acknowledges hospitality of Case 
Western Reverse University, where part of this work was done. This work was supported by the Friedrich-
Ebert-Foundation (M.S) and Deutsche Forschungsgemeinschaft grant IRTG 881 (M.S. and D.J.S.) and by a 
grant from the US-DOE to the particle astrophysics theory group at CWRU.

\end{document}